# A Hazard Analysis Technique for Additive Manufacturing

Gregory Pope[1], Mark Yampolskiy[2]

[1]Lawrence Livermore National Laboratory, [2]University of South Alabama

Additive manufacturing conceptually consists of four basic steps that allow a Computer Aided Design (CAD) to be transformed into a completed physical part. The simplified diagram below illustrates these four basic steps. A 3D CAD model is decomposed into horizontal slices by slicing software and then printed layer by layer using a 3D printer.

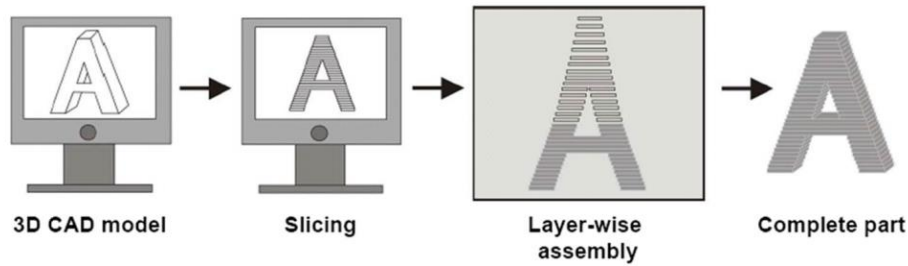

The promise of Additive Manufacturing (AM) includes reduced transportation and warehousing costs, reduction of source material waste, and reduced environmental impact[1]. AM is extremely useful for making prototypes and has demonstrated the ability to manufacture complex parts not possible (or prohibitively expensive) with conventional machining. Scientists and manufactures are finding increased uses for AM in creation of all types of finished products including those built from polymers, biological material, and metals. Although companies such as GE have been using 3D printing for Additive Manufacturing for over thirty years to make mandrels for light bulb manufacturing[2], application areas of Additive Manufacturing have increased substantially in recent years, particularly due to the reduction in cost of 3D printers. Wohler's Associates reported a compound annual growth rate (CAGR) for AM of 35.2% to $4.1 billion in 2014.[3] AM is predicted to continue to grow rapidly for the next four years as indicated in the graph below[4].

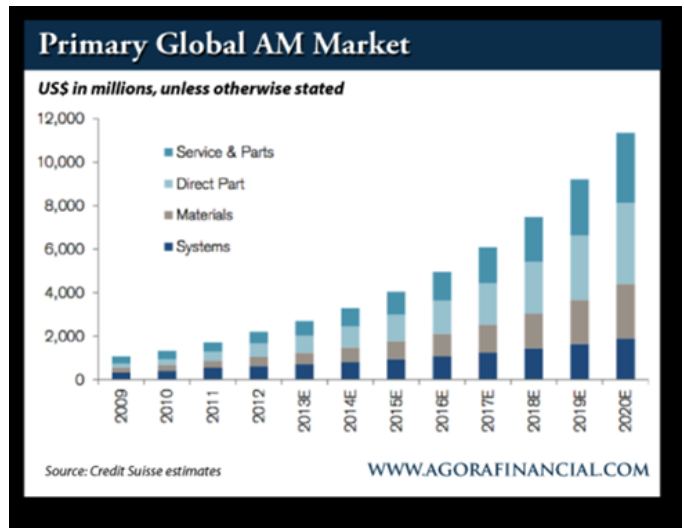

Like most emergent technologies, there are bound to be growing pains with AM. This paper looks at the software that supports AM and 3D printing and their vulnerability to cyber-attacks, intellectual property theft, defect rates of AM software (which can cause undesired consequences themselves and also create vulnerabilities that a hacker may exploit), part reliability and safety of devices incorporating 3D printed parts (when making mission critical parts), and security/throughput issues of computer networks. Literature searches, consulting with technical experts and a relatively new hazard analysis technique will be used, one especially developed for software intensive systems called Systemic Theoretic Process Analysis (STPA)[5]. The purpose of this white paper is to identify risks (or hazards for mission critical parts) for AM in this emergent stage so that mitigations can be applied before accidents occur. A second purpose of this white paper is to evaluate the effectiveness of STPA as a hazard analysis technique in a field that is still relatively new.

There are two main user groups for 3D printing, the hobbyists or desktop 3D printer users who can add an interface card to their PC and produce novelty items on a 3D printer costing less than $5,000, and the industrial users who use larger high quality 3D printers to make prototypes and functional parts using a variety of materials, most commonly polymers and metals. GE has test fired a mini 3D printed turbine engine.[6]. GE has also delivered two of their new LEAP-1A jet engines, which has additively manufactured fuel nozzles[7][8]. These industrial grade 3D printers are cost anywhere from $30,000 to $2,000,000. The promise of printing out spare parts only when needed at almost any internet connected location worldwide has attractive benefits for logistics support.

Currently ASTM has standardized on seven different 3D printer technologies: [9]

1. **Binder jetting**, an additive manufacturing process in which a liquid bonding agent is selectively deposited to join powder materials.
2. **Directed energy deposition**, an additive manufacturing process in which focused thermal energy is used to fuse materials by melting as they are being deposited.
    Discussion —"Focused thermal energy" means that an energy source (e.g., laser, electron beam, or plasma arc) is focused to melt the materials being deposited.

3. **Material extrusion**, an additive manufacturing process in which material is selectively dispensed through a nozzle or orifice.
4. **Material jetting**, an additive manufacturing process in which droplets of build material are selectively deposited.
   Discussion—Example materials include photopolymer and wax.
5. **Powder bed fusion**, an additive manufacturing process in which thermal energy selectively fuses regions of a powder bed.
6. **Sheet lamination**, an additive manufacturing process in which sheets of material are bonded to form an object.
7. **Vat photopolymerization**, an additive manufacturing process in which liquid photopolymer in a vat is selectively cured by light-activated polymerization.

The following table lists examples of materials used for the seven different 3D printing techniques:

| 3D Printer Technology | Materials Used |
|---|---|
| **Binder jetting** | Powdered material: Metals- stainless steel, copper Polymers: ABS, PA, PC, Ceramics- glass and binder material- liquid bonding agents |
| **Directed energy deposition** | Cobalt Chrome, Titanium |
| **Material extrusion** | Thermoplastics, eutectic metals, edible materials, rubbers, modeling clay, plasticine, metal clay (including precious metal clay) |
| **Material jetting** | Ceramic materials, metal alloy, cermet, metal matrix composite, ceramic matrix composite |
| **Powder bed fusion** | Almost any metal alloy, powdered polymers, plaster, cobalt chrome alloys, stainless steel, aluminum, thermoplastics, metal powders, ceramic powders |
| **Sheet lamination** | Paper, metal foil, plastic film |
| **Vat photopolymerization** | Liquid photopolymer resin |

The AM process normally requires data transmission from the CAD/CAM system and repository to the 3D printer. To support this requirement various file formats are used. The internal CAD representations of a three dimensional object is sent in either ASCII or Binary STL (stereolithographic) format which describes objects using triangle vertices and their normals. A short example of ASCII STL is shown below[10]:

```
cube_ascii.stl (C:\TEMP) - GVIM
File  Edit  Tools  Syntax  Buffers  Window  Help

solid ascii
  facet normal 9.461808e-017 -0.000000e+000 1.000000e+000
    outer loop
      vertex   1.443618e+000 -5.518407e+000 1.280603e+000
      vertex   1.443618e+000  0.000000e+000 1.280603e+000
      vertex  -1.443618e+000  0.000000e+000 1.280603e+000
    endloop
  endfacet
  facet normal 9.461808e-017 0.000000e+000 1.000000e+000
    outer loop
      vertex  -1.443618e+000  0.000000e+000 1.280603e+000
      vertex  -1.443618e+000 -5.518407e+000 1.280603e+000
      vertex   1.443618e+000 -5.518407e+000 1.280603e+000
    endloop
  endfacet
  facet normal 1.000000e+000 0.000000e+000 -1.066625e-016
    outer loop
      vertex   1.443618e+000 -5.518407e+000 -1.280603e+000
      vertex   1.443618e+000  0.000000e+000 -1.280603e+000
      vertex   1.443618e+000  0.000000e+000  1.280603e+000
    endloop
  endfacet
  facet normal 1.000000e+000 0.000000e+000 -1.066625e-016
```

While being a de facto standard, STL describes the surface of an object; it does not contain information about the objects texture, color, material, substructures, and other properties. To address this, additional file formats have evolved such as AMF[11] and OBJ.

3D printers receive files that describe the object to be printed in different ways

1. Individual commands to 3D printer (G-Code) [12]

2. Complete CAD file transferred (STL/AMF or proprietary formats)

3. Some sort of mixture – a single "toolpath" file with AM machine specific instructions

These files must travel from a CAD/CAM presumably to some type of a repository and then to a 3D printer. These files can represent the intellectual property of the enterprise and as such need to be carefully protected when stored or transmitted.

There are a number of open source offerings and nearly 20 commercial companies who provide 3D software products for both beginner, intermediate, and advanced users[13]. The quality of this relatively new software must be sufficient to protect intellectual property, be immune to cyber-attack, and support error free creation of parts, especially for mission critical parts for industries such as transportation, medical, and defense.

Before engaging in the use of STPA for Hazard Analysis of the AM process, a search was conducted to review literature already existing in the area of AM software. One white paper was found from the University of Alabama Huntsville titled "Additive Manufacturing Security[14]" which discussed seven different attack vectors available to hackers and the potential consequences of these attacks. A second source of information was found in Chapter 11 of IFIP working group 11.10 on Critical Infrastructure Protection book titled: "Security Challenges with Additive Manufacturing with Metals and Alloys" [15] which discussed five different attack vectors and their impact on 14 different manufacturing parameters. Further searching identified a number of publications by Mark Yampolskiy et al. from the University of Southern Alabama with additional concerns about AM security. From these excellent reference documents, the need to protect AM from a wide variety of attacks from various sources is apparent. The question was then, could STPA identify any additional security concerns that have not been discussed in the existing literature?

STPA is a technique that is based on viewing the entity to be analyzed as a system. In its most basic form STPA would view a control system as shown below:

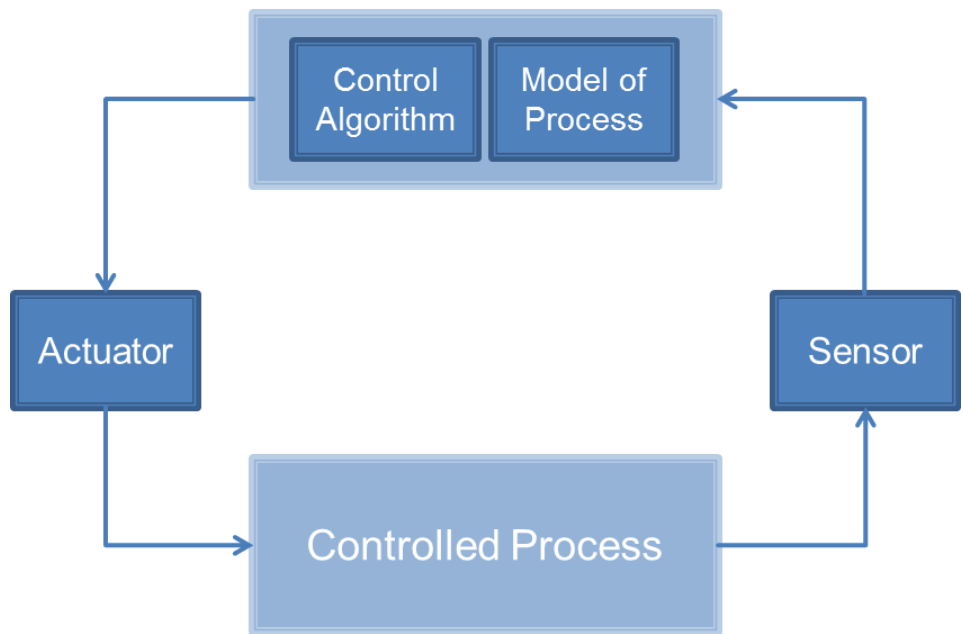

In this simplified representation of a real time control safety system, a software controller is shown consisting of a control algorithm and model of the process being controlled. The controller receives information from a sensor and based on that information may initiate control through an actuator. After creating the system model, guide phrases are used on the paths between components to detect hazards. Hazards are defined as an undesirable outcome that could lead to an accident. The guide phrases are shown below for the actuator path for a real time safety control system:

1. A control action required for safety is not provided or is not followed.

2. An unsafe control action is provided that leads to a hazard.

3. A potentially safe control action is provided too late, or out of sequence.

4. A safe control action is stopped too soon or applied too long.

The guide phrases also apply to the sensor path:

1. A sensor reading required for safety is not provided or is not followed.

2. A sensor reading is provided that leads to a hazard.

3. A sensor reading is provided too late, or out of sequence.

4. A sensor reading is stopped too soon or applied too long.

Hazards may occur because a software algorithm is incorrect or the model of the process in the software controller and the actual process are different. An example of an incorrect algorithm accident would be the Arian 5 rocket failure on June 4, 1996. The software algorithm used to measure horizontal velocity to correct attitude used in the Arian 5 rocket was reused from the Arian 4 rocket. The more powerful Arian 5 rocket was capable of more horizontal velocity than the Arian 4. The Arian 4 algorithm was never tested beyond the maximum horizontal velocity of the Arian 4 rocket. The Arian 5's flight path was considerably different and beyond the range for which the reused computer program had been designed. The 64 bit floating point horizontal velocity at 37 seconds after launch was too large to be converted into a 16 bit integer which the attitude control algorithm used in both the primary and back up inertial guidance computers. An arithmetic overflow error then occurred and spurious data was sent to the rocket's nozzles causing the rocket go out of control[16]. The Arian 5 accident shows how reused software may not work as expected, that hazards can occur without the software failing, and switching to a backup version of the same software does not help. In this accident billions of dollars of satellites were destroyed. For AM software a similar kind of software system issue could lead to faulty automotive parts causing the recall of millions of cars, grounding of aircraft, loss of reputation, and litigation.

An example of controller model and the controlled process being different accident would be the Mars Polar lander on December 3, 1999 after an 11 month transit from earth. The software on board the Mars Polar lander was designed to fire a retrorocket to slow the lander's vertical descent prior to touching down. The retrorocket would be commanded to stop firing when the computer sensed that one of the three Mars Polar lander's legs touched down on the Martian surface. The algorithm was sensing the leg touch down circuit to determine the spacecraft had landed. The decent to the Martian surface proceeded as planned, 70 to 100 seconds before touch down the landing legs deployed and the retrorockets began warming up. At about 1.4 km altitude the retrorockets were to begin firing to slow the spacecraft for landing. However it is speculated that the mechanical deployment of the lander legs at speed caused a transient spike in the touch down circuit that made the computer think that a leg had touched down on the surface. The computer then turned off the retrorockets as it was required to do, sending the lander crashing into the surface of Mars at a high rate of speed[17].

STPA proceeds using the guide phrases on the various interconnecting paths. STPA focuses on the interactions between components which are the most common failure modes in software based systems. STPA attempts to uncover the types of conditions that caused the Arian 5 and Mars Polar lander accidents before they happen.

A human operator or user is also a common component of a system, shown below the expanded STPA model opens up additional interaction paths to apply the guide phrases:

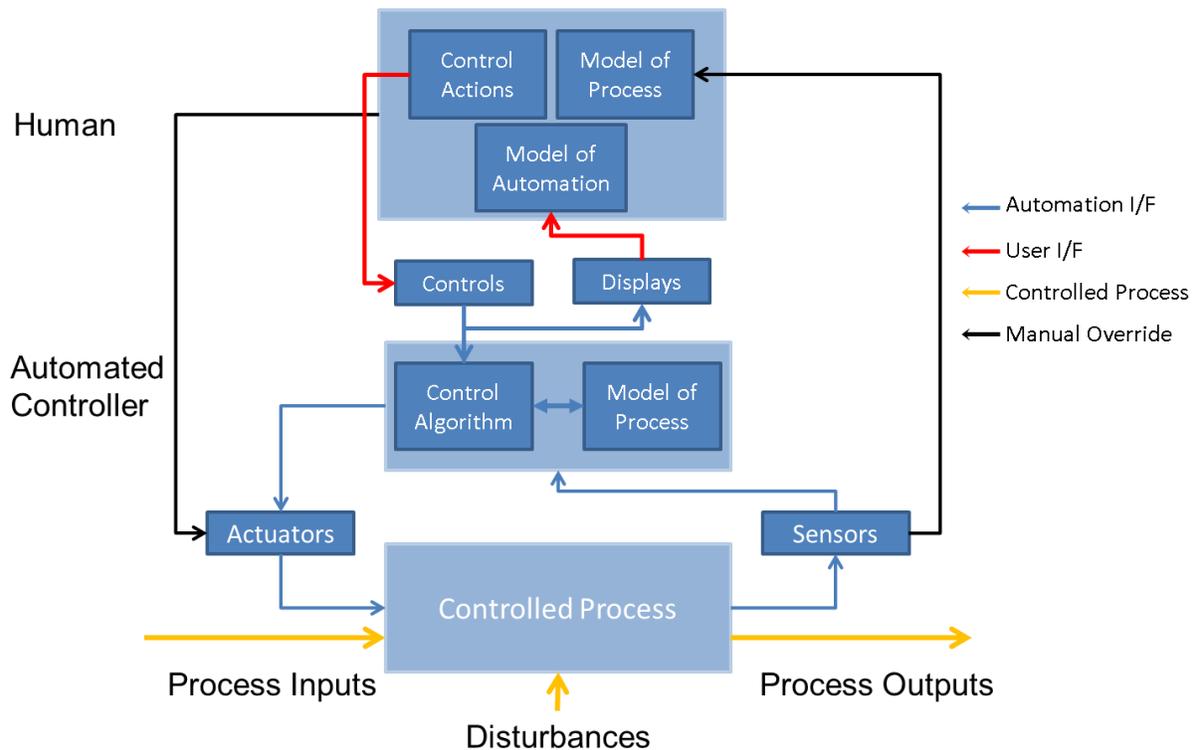

The human in the system also has a mental model of the controlled process and a mental model of the automation controlling the process. The human commonly interacts with controls and displays which create more interconnected paths and additional potential hazards.

An example of a mental model accident was the Asiana 214 crash on July 6th, 2013 at the end of runway 28L at San Francisco International Airport. The pilot in command was sure the Boeing fly-by-wire flight automation of the 777 would prevent the airspeed from decreasing to an unsafe value (below stall speed) while landing. The automated system worked exactly as designed and in the mode it was in assumed the pilot was controlling the throttles, which the pilot had previously set to idle. The pilot's mental model of how the automation worked was incorrect. It is interesting to note that the pilot in command's prior experience with fly-by-wire automation was in Airbus aircraft. The Airbus fly-by-wire automation philosophy does override pilot inputs that may be harmful, whereas Boeing's philosophy is to allow the pilot to override the automation. The differing approaches to aircraft automation by Airbus and Boeing aircraft may have been a contributing factor to the pilot's flawed mental model of the automation[18].

Controls can also create accidents as demonstrated in the TransAsia 235 crash after takeoff from Taipei Songshan airport on February 4, 2015. Just 37 seconds after takeoff the right engine (engine 2) on the ATR 72-600 turbo prop indicated failure. The pilot then cut the power to the left engine (engine 1), instead of the right engine. The aircraft was then without any engine power setting the stage for the crash to follow. The ATR 72-600 aircraft has two throttles, number 1 and 2. The throttle handle for engine 1 is twice as big as the throttle handle for engine 2. By the time the pilot realized he had cut power to the working engine it was too late to recover. The larger size of the left engines throttle handle may have been more instinctual to pull back in an emergency[19].

Notice that in the accident examples cited above the software was doing what it was designed to do, the hazards were in the interaction between the software and other system components or humans. STPA attempts to find these types of potential hazards before accidents occur. Or in the case of non-safety related systems identify risks or security vulnerabilities that could cause undesired consequences and mitigate them.

Non-real-time processes can also benefit from STPA. The AM process is depicted as a system as shown below:

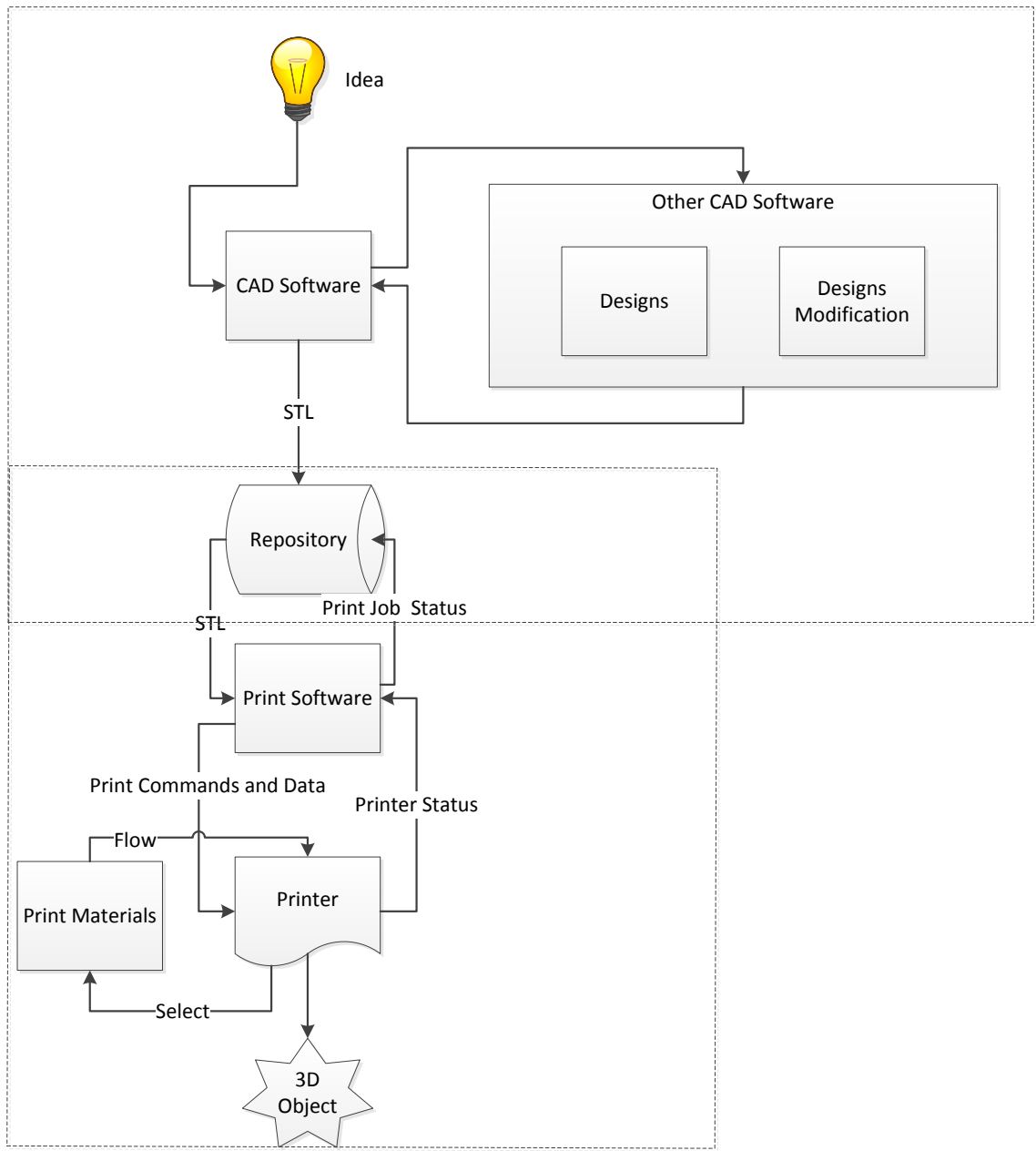

The AM system is analyzed as three subsystems, the CAD/CAM sub-system, the repository subsystem, and the 3D printing subsystem. In the diagram below the guide phrases modified for non-real time are applied to the CAD/CAM subsystem:

STPA Guide Phrases

A resource or action required for correct operation is not provided or is not followed.

An incorrect resource or action is provided that leads to a hazard/risk.

A potentially correct resource or action is provided too late, or out of sequence.

A correct resource or control action is stopped too soon or applied too long.

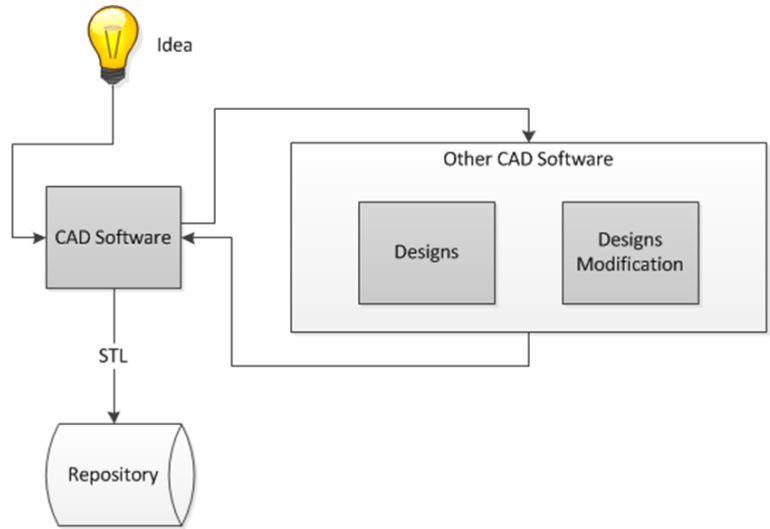

STPA is useful as a risk analysis tool when analyzing undesired consequences which may not cause physical injury. STPA analysis proceeds by applying the guide phrases above to each interface and component of the CAD/CAM subsystem. The process is repeated for each sub-system:

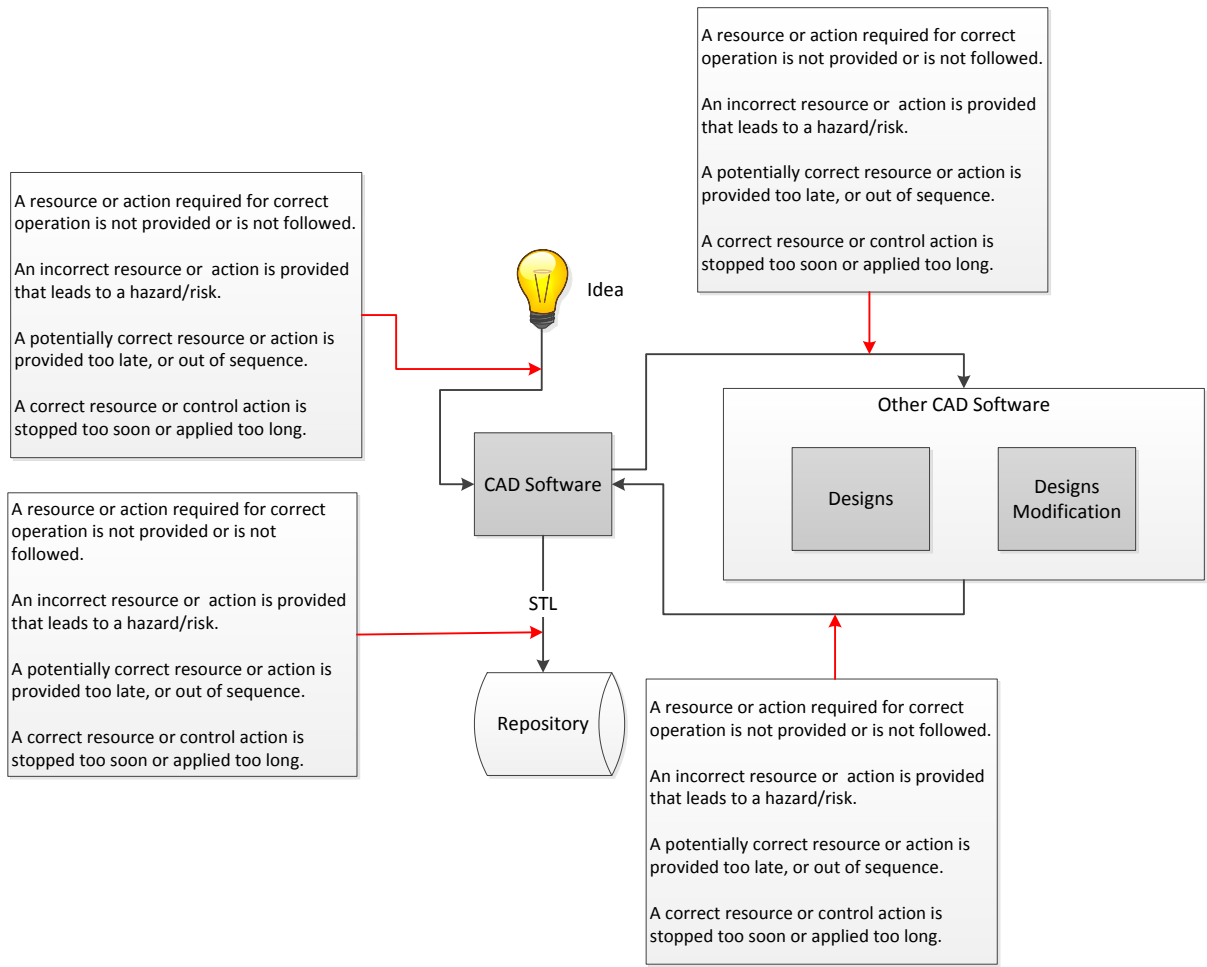

Asking the four guide phrases for each interface and component yielded the following list of concerns for the CAD/CAM subsystem:

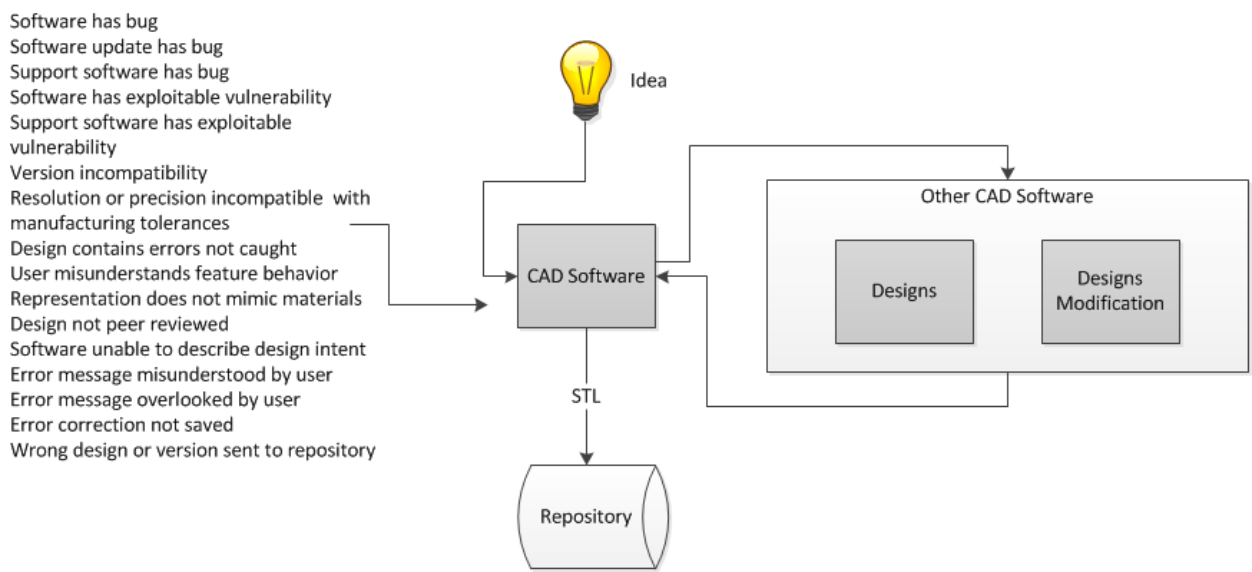

Next the repository subsystem was evaluated using the same guide phrases and the following potential risks or hazards were found:

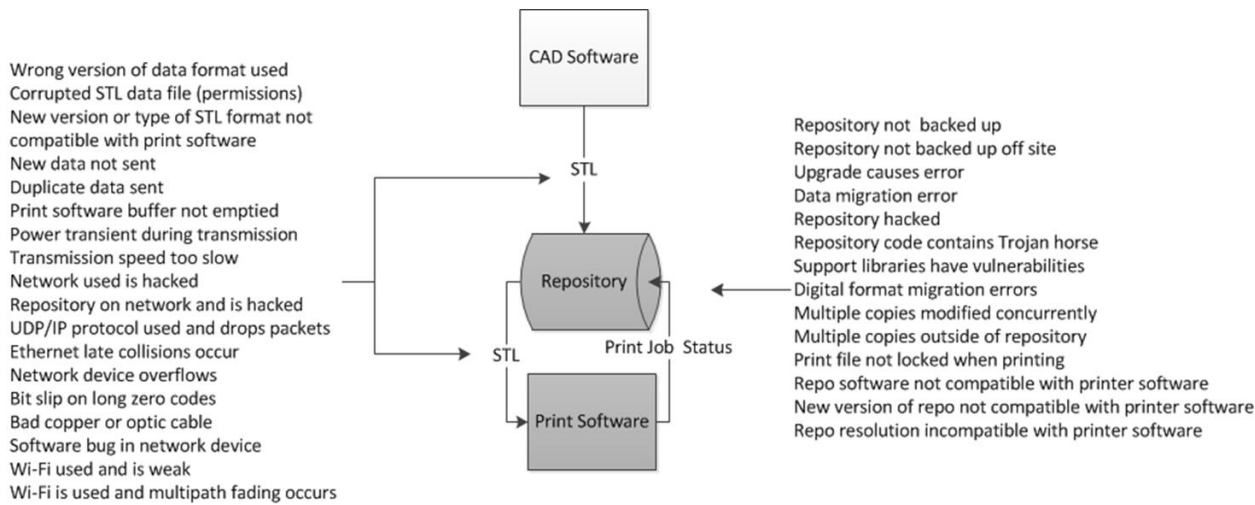

The same guide phrases were applied to the 3D printing subsystem and a number of potential risks or hazards were identified:

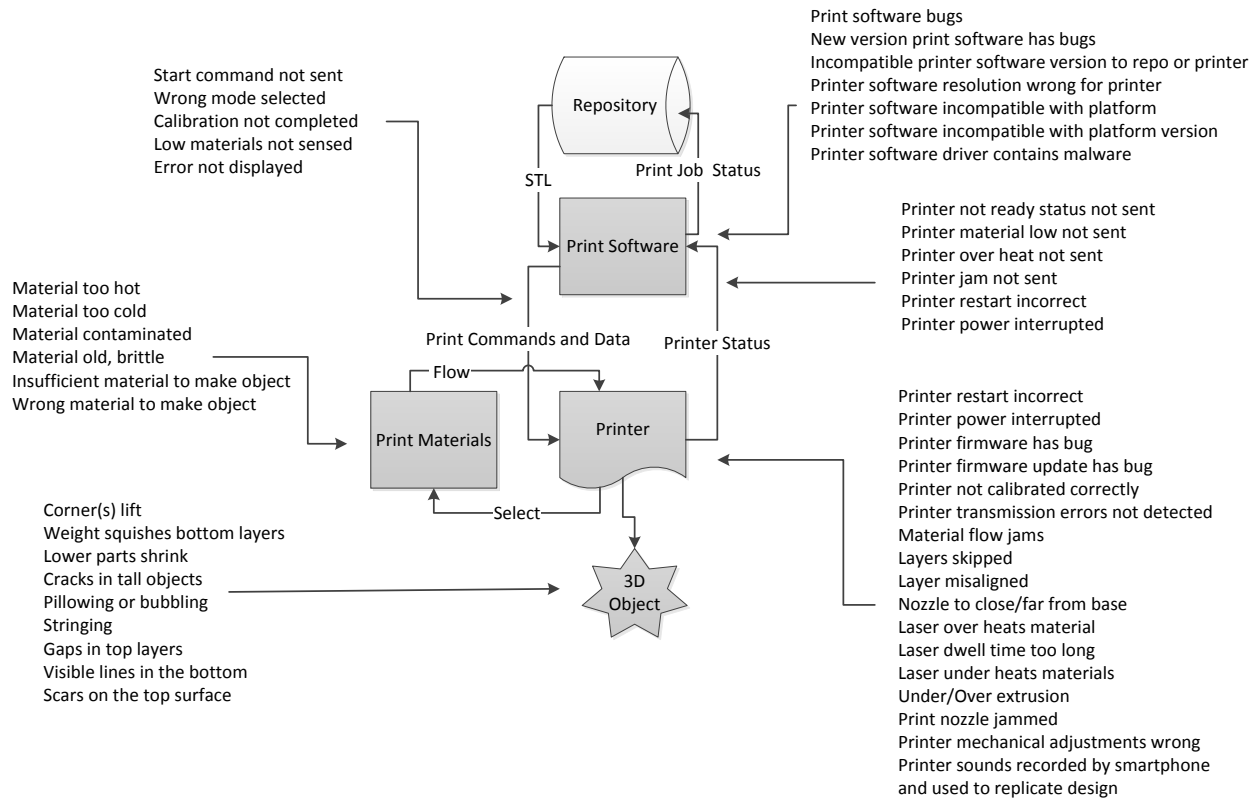

In addition to preforming STPA to identify potential AM risks (which become hazards if creating mission critical parts), an additional software reliability study was done to determine the defect rate of software used for Factory Automation. Software reliability can be measured in terms of defects in the code. Defects in the AM software can cause undesired consequences themselves and also create vulnerabilities that a hacker may exploit. Software reliability generally improves as a version of a software product is widely used and problems found with it are repaired or patched. Software tends to be less reliable when it is initially deployed. Studies have been conducted over the years for different types of software, using errors per thousand lines of code (KSLOC). In order to compare software written in different languages the unit of measure is KESLOC (thousands of equivalent source lines of code) normalizing for language differences. The following industrywide table compiled in 2004[20] by Donald Reifer[21] indicates KESLOC values for newly released code:

| Application Domain | Number Projects | Error Range (Errors/KESLOC) | Normative Error Rate (Errors/KESLOC) | Notes |
|---|---|---|---|---|
| Automation | 55 | 2 to 8 | 5 | Factory automation |
| Banking | 30 | 3 to 10 | 6 | Loan processing, ATM |
| Command & Control | 45 | 0.5 to 5 | 1 | Command centers |
| Data Processing | 35 | 2 to 14 | 8 | DB-intensive systems |
| Environment/Tools | 75 | 5 to 12 | 8 | CASE, compilers, etc. |
| Military -All | 125 | 0.2 to 3 | < 1.0 | See subcategories |
| ▪ Airborne | 40 | 0.2 to 1.3 | 0.5 | Embedded sensors |
| ▪ Ground | 52 | 0.5 to 4 | 0.8 | Combat center |
| ▪ Missile | 15 | 0.3 to 1.5 | 0.5 | GNC system |
| ▪ Space | 18 | 0.2 to 0.8 | 0.4 | Attitude control system |
| Scientific | 35 | 0.9 to 5 | 2 | Seismic processing |
| Telecommunications | 50 | 3 to 12 | 6 | Digital switches |
| Test | 35 | 3 to 15 | 7 | Test equipment, devices |
| Trainers/Simulations | 25 | 2 to 11 | 6 | Virtual reality simulator |
| Web Business | 65 | 4 to 18 | 11 | Client/server sites |
| Other | 25 | 2 to 15 | 7 | All others |

Notice that Factory Automation code, the class of code AM software would belong to have a normative error rate of 5 KESLOC. Mission critical codes, such as those used for avionics have normative error rates about an order of magnitude lower than Factory Automation software. This makes sense since in theory mission critical codes are much more carefully designed and tested. However if we are going to be using Factory Automation software to make mission critical parts for AM, such as components of a commercial gas turbine engine, or an automobile, or missile we would prefer to see the software used to make those parts have lower defect rates.

Defects in the software code can be viewed as structural or functional. Structural defects (memory leaks, not initializing a variable prior to use, overflowing an array bounds, not checking input data before use in a loop index or pointer, inputs depending on user supplied delimiters, error conditions not handled, etc.) create exploitable vulnerabilities for hackers. About 38% of defects are of this type and easily detectable with static analyzers such as Klocwork or Coverity. Often structural defects cause intermittent problems that are difficult to detect with testing. The other major type of defect category is the functional type. The software code does not implement the software requirements correctly. These types of defects are detected using unit testing and functional test cases at the integration and full system test level. Functional testing coverage can be measured using dynamic analyzers such as Gcov or Intel's compiler coverage feature. Negative testing should also be included to protect against hackers trying to deliberately insert SQL statements or extremely large files into data fields.

A more recent defect number for Factory Automation software defect rates was obtained from Donald Reifer, measured at one year after release to give the software a chance to mature and fix additional defects found by users. The 2016 number for fault density in Factory Automation software is 2 per KESLOC, still four times more than newly released mission critical software. The concern is much of the AM software; especially slicer software and 3D printer firmware are relatively new and may not be at a high enough maturity level to be safely used for manufacturing mission critical parts or are well protected from hackers.

As a result of using STPA for AM a number of risks (or hazards, depending on what is being printed) emerged that had not been found in the literature or thought of intuitively. The discovery of these risks/hazards prompted additional research into specific topic areas such as network level protocols and error handling. STPA also helped produce good questions to ask of experts in various specialty areas. It also produced good ideas for test cases, where errors can be seeded into files sent to 3D printers to assure they are detected.

Identifying the risks is the first step in a risk management program. Next for identified risks in this study an example set of mitigations was listed below. Many of the STPA identified AM risks/hazards discovered in this analysis can be mitigated by:

1. Assuring the network protocol used for AM is TCP/IP and not UDP which does not guarantee error free transmissions. UDP is commonly used for voice over IP and video transmissions.
2. Assuring 3D printer buffer size is large enough to hold the entire print image so that error free receipt of a print image can be assured before printing starts. Otherwise a part may have to be scrapped before it is completely printed if there is a transmission error occurring in later transmissions.
3. Assuring that networks used for AM have a high Quality of Service [delay, delay variation (jitter), bandwidth, and packet loss parameters]. Dropped packets could slow the printing process consequently cause the quality degradation of the printed part.
4. Assuring that 3D printer software can detect a transmission error from the network software so it does not attempt to print data that is corrupted.
5. Assuring application level data has an integrity check (EDC/ECC codes, word count).
6. Assuring repositories containing design data are encrypted and backed up offsite automatically on a regular basis.
7. Assure repositories tightly control access to part files and provide file configuration management.
8. Assuring any transmission of design data is encrypted.
9. Assuring 3D software and supporting software, especially open source software, has been checked for Trojans, worms, viruses, and other types of malware.
10. Dedicating enterprise networks used for AM and air-gap them to the internet.

11. Being cautious about using any commercial 3D printing software that has not had time to mature with a wide distribution of users over many months. Refrain from being an early adopter of new software used for production parts. This applies to updates as well.
12. Assuring that AM software has been subjected to static code analysis to identify and remove structural errors that hackers could exploit.
13. Assuring AM software and supporting libraries have been scanned for known vulnerabilities.
14. Assuring AM software has only verified I/O operations that are intentional.
15. Assuring AM software has been subjected to dynamic code analysis to measure test coverage and memory management.
16. Choosing fiber optic physical networks for AM over copper cables or wifi because of fiber's immunity to EMI/EMC and radio wave interference.
17. Conducting an end to end (CAD/CAM, Repository, Slicers, 3D printer Software, data formats) system analysis of AM system components to assure that ranges, resolutions, accuracies, engineering units, and formatting options are compatible and adequate.
18. Assuring adequate training has been conducted for users of the AM system (CAD/CAM, Repository, 3D Printers and Software)
19. Assuring software upgrades are evaluated on a test platform before committing to a production system.
20. Developing an end to end system test object that can be printed and verified prior to using the AM system for a production run.
21. Assuring calibration and mechanical alignment of the 3D printer is conducted on recommended intervals or more frequently if required.
22. Keeping audio recording devices, including cell phones, out of the 3D printer area, some 3D printer mechanical mechanisms generate acoustical noise that is unique for each printing action and can be reproduced if recorded.
23. Keeping appropriate fire suppression equipment in close proximity to the printer area.
24. Assuring transient suppression and auxiliary or battery power back up exists for 3D printer.
25. Assuring 3D printer power can be shut off from a master power switch a safe distance from the printer.

This list of mitigations is only a partial set and may not apply to every AM installation. The list serves as an example of how to use STPA to identify risks to create mitigations. Mitigations were created by going through the list of hazards and risks and asking what action could be taken to prevent or minimize the hazard or risk. Hopefully this paper will acquaint the reader with how to use STPA on their actual AM enterprise configurations and discover their hazards and risks and take mitigation steps before accidents or losses occur.

In conclusion, STPA combined with a literature search and asking the experts questions, was definitely useful in determining hazards, risks, and security vulnerabilities not intuitively obvious. STPA was useful in formulating questions for experts in networks and cyber security domains. Experts were most useful in confirming which mitigations already exist in the AM system model and which should be added in the future.

End Notes

[16] https://www.ima.umn.edu/~arnold/disasters/ariane5rep.html
[17] http://spaceflight.nasa.gov/spacenews/releases/2000/mpl/mpl_report_1.pdf
[18] http://www.ntsb.gov/news/events/Pages/2014_Asiana_BMG.aspx
[19] https://en.wikipedia.org/wiki/TransAsia_Airways_Flight_235
[20] Because of the date of this study decided to also directly contact Donald Reifer to get the latest defect rates for Factory Automation software to be revealed in the following paragraph. The 2016 defect rate is lower, but also is measured one year after deployment instead of at release which was the method used in 2004.
[21] Donald Reifer, "Industry Software Cost, Quality, and Productivity Benchmarks", DoD Software Tech News, July 2004